\documentstyle[prl,aps,multicol]{revtex}

\begin{document}
\newcommand{\ket}[1] {\mbox{$ \vert #1 \rangle $}}
\newcommand{\bra}[1] {\mbox{$ \langle #1 \vert $}}
\newcommand{\bkn}[1] {\mbox{$ < #1 > $}}
\newcommand{\bk}[1] {\mbox{$ \langle #1 \rangle $}}
\newcommand{\scal}[2]{\mbox{$ < #1 \vert #2 > $}}
\newcommand{\expect}[3] {\mbox{$ \bra{#1} #2 \ket{#3} $}}
\newcommand{\ki}{\mbox{$ \ket{\psi_i} $}}
\newcommand{\bi}{\mbox{$ \bra{\psi_i} $}}
\newcommand{\p} \prime
\newcommand{\e} \epsilon
\newcommand{\la} \lambda
\newcommand{\om} \omega   \newcommand{\Om} \Omega
\newcommand{\cc}{\mbox{$\cal C $}}
\newcommand{\w} {\hbox{ weak }}
\newcommand{\al} \alpha
\newcommand{\bt} \beta
\newcommand{\be} {\begin{equation}}
\newcommand{\ee} {\end{equation}}
\newcommand{\ba} {\begin{eqnarray}}
\newcommand{\ea} {\end{eqnarray}}
\def\lrD{\mathrel{{\cal D}\kern-1.em\raise1.75ex\hbox{$\leftrightarrow$}}}
\def\lr #1{\mathrel{#1\kern-1.25em\raise1.75ex\hbox{$\leftrightarrow$}}}
\title{ Quantum metric fluctuations and Hawking radiation }
\author{R. Parentani
}
\address{
Laboratoire de Math\'{e}matiques et Physique
Th\'{e}orique, 
CNRS UPRES A 6083,
Universit\'{e} de Tours, 
37200 Tours, France}

\maketitle

\begin{abstract}
In this Letter we study the gravitational interactions
between outgoing configurations giving rise to Hawking radiation
and in-falling configurations. When the latter are in their ground state, 
the near horizon interactions lead to collective effects 
which express themselves as metric fluctuations and
which induce dissipation, as in Brownian motion. 
This dissipation prevents the appearance of trans-Planckian 
frequencies and leads to a description of Hawking radiation
which is very similar to that obtained  from sound propagation
in condensed matter models.

\vskip .3 truecm

\noindent
PACS number(s): 04.70.Dy, 04.60.-m, 05.40.-a



\end{abstract}


\begin{multicols}{2}


In his original derivation\cite{Hawk}, Hawking assumed
that gravity can be treated classically, 
i.e. the metric was determined by the energy of the collapsing star
but was unaffected by the quantum processes under examination.
In this approximation, the radiation field
satisfies a linear equation (in the absence of matter interactions)
and the in-falling and outgoing field 
configurations are completely uncorrelated near the black hole horizon.
In fact the pairs of quanta generated by its formation
are composed of two outgoing quanta, one of each side of it.
The external ones form the asymptotic flux whereas
their partners fall towards the singularity at $r=0$.
Upon tracing over these inner configurations 
one gets an outgoing incoherent flux described by a thermal density matrix. 

There is nevertheless a precise relationship between the 
expectation values of the in-falling and the outgoing energy fluxes. 
Indeed, the asymptotic thermal flux 
is accompanied by a negative in-falling flux $\langle T_{vv}\rangle$
which has, on  the horizon, exactly the opposite value and
which drives black hole evaporation according to 
\be
{ dM \over dv } = \langle T_{vv}\rangle \vert_{r=r_{horizon}=2M} 
\simeq -{  1 \over M^2}\, .
\label{one}
\ee
We work with in Planck units: $c=\hbar =  M_{Planck}=1$.
$v= t + r^*$ is the advanced null time,
$r^*= r + 2M \ln (r /2M -1 )$ is the tortoise coordinate
and $M(v)$ is the time dependent mass which appears in the 
Vaidya metric 
\be
ds^2 = - ( 1 - { 2M \over r}) dv^2  + 2 dv dr + r^2(d\theta^2 + sin^2 \theta d\phi^2) 
\label{two}
\ee
The important point is that (\ref{two})
offers a good approximation of the near horizon geometry
(i.e. $\vert r- 2M\vert \ll 2M$)
of an evaporating black hole, see \cite{massar,GO}.

This semi-classical description 
would be perfectly valid if another feature of black hole physics
wasn't present, namely 
the field configurations giving rise to Hawking quanta
possess arbitrary high (trans-Planckian) frequencies near the horizon:
When measured by in-falling observers at $r$,
the frequency grows as
\be
\om = {\la \over 1 - 2M /r } \simeq { \la \, 2M\over r - 2M} 
\label{dopef}
\ee
where $\la$ is the asymptotic energy of the quantum.
This implies that a wave packet centered along the 
null outgoing geodesic $u=t-r^*$ had a frequency 
$\om=\la  e^{\kappa u}$ when it emerged from the in-falling star.
($\kappa=1/4M$ is the surface gravity and fixes Hawking temperature 
$T_{H}= \kappa/ 2 \pi$.)
Unlike processes characterized by a typical energy scale, 
the relation $\om=\la  e^{\kappa u}$ shows that black hole
evaporation rests on arbitrary high frequencies. 

As emphasized by 't Hooft\cite{THooft}, this implies that  
the gravitational interactions between these configurations 
and in-falling quanta cannot be neglected, thereby questioning the validity of 
the semi-classical description.
In questioning this validity, two issues 
should be distinguished, see Sect. 3.7 in \cite{GO}.
First, there is the question of the low frequency $O(\kappa)$ changes 
which can be measured asymptotically, and secondly, that 
of the high frequency modifications of the near horizon physics.
It should also be stressed that thermo-dynamical reasonings indicate that 
the asymptotic properties, namely thermality governed by $\kappa$ 
and stationarity, should be preserved.
The difficulty is therefore to conciliate the stability of these properties 
with the radical change of the near horizon physics
which is needed to cure the trans-Planckian problem.
Indeed, local interactions near the horizon lead to recoil
effects proportional to $\om$ which seem incompatible 
with the stationarity of the flux\cite{hawkfr}.
It thus appears inappropriate to treat these interactions
by a perturbative approach based on Feynman diagrams. 

A different approach is suggested
by the analogy with condensed matter physics 
pointed out by Unruh\cite{Unruh81}.  
He noticed that sound propagation in a moving fluid obeys a d'Alembertian 
equation which defines an acoustic metric. Therefore, upon neglecting 
viscosity and non-linear effects, thermally distributed photons 
should be emitted when the acoustic metric corresponds to that 
of a collapsing star. (The formation of an acoustic horizon
occurs when the fluid velocity reaches the speed of sound.)
However contrary to photons, 
the dispersion relation of phonons is not
linear for frequencies (measured in the rest frame of the fluid) 
higher that a critical $\om_c$. 
Nevertheless, when $\om_c \gg \kappa$,
the asymptotic properties of the Hawking flux of phonons are 
unaffected \cite{dumbunruh}
in spite of the fact that the high frequency spectrum, 
which was solicited in Hawking's derivation, is no longer available.

This insensitivity suggests that something similar might apply 
to black holes and solve the trans-Planckian problem.
However it is a priori unclear what can play the role of the microscopic 
constituents of the fluid which introduce through their interactions 
the non-trivial dispersion relation and the cut-off $\om_c$ 
which breaks the low frequency Lorentz invariance\cite{jaco}.

The aim of this letter is to show that the gravitational interactions 
between the outgoing configurations giving rise to Hawking radiation
and in-falling configurations in their ground state lead to collective effects 
which define an effective dispersion relation for the outgoing modes.
These collective effects express themselves in terms of 
a stochastic ensemble of metric fluctuations.
The specification of the vacuum state at early times
determines the statistical properties of this ensemble and this 
in turn fixes both the cut-off $\om_c$ (in terms of $\kappa$)
and the frame which breaks the 2D Lorentz invariance\cite{2dhili}.


For simplicity, we shall consider only s-waves
propagating in spherically symmetrical space times.
The background metric is taken to be that generated 
by the collapse of a null spherical shell of mass $M_0$ which propagate along 
$v=0$. Inside the shell, for $v<0$, the geometry is Minkowski
and described by (\ref{two}) with $M=0$. Outside, the metric is also static and
given by (\ref{two}) with $M=M_0$. 

When studying the propagation of massless s-waves in this background,
they fall into two classes according to their support on ${\cal{J}}^-$, 
the light-like past infinity.
The waves in the first class have support only for 
$v<0$ and will be noted  $\phi_-$. They propagate inward in the flat geometry 
till $r=0$ where they bounce off and become outgoing configurations. 
This first class is itself divided in two: For $v<-4M$, 
the reflected waves cross the in-falling shell with 
$r>2M$ and reach the asymptotic region\cite{f1}
whereas those for $0>v>-4M$ cross it with $r<2M$ and
propagate in the trapped region till the singularity.
The separating light ray $v_H=-4M$ becomes the future horizon $u=\infty$
after bouncing off at $r=0$. 
The configurations which form the second class have support only for $v>0$ 
and are noted $\phi_+$. They are
always in-falling and cross the horizon towards the singularity. 

In the semi-classical derivation of black hole radiation, the configurations for $v<v_H$
give rise to the asymptotic quanta, those for $v_H<v<0$ to their partners
whereas $\phi_+$ 
plays no role in the asymptotic radiation.
The correlations between the asymptotic quanta and their partners
follow from the fact that, on ${\cal{J}}^-$ and in the vacuum,
the rescaled field $\phi = r \chi$ (where $\chi$ is the 4D s-wave) satisfies 
\be
\langle \phi(v) \phi(v') \rangle = \int_0^{\infty}{ d\om \over 4\pi \om } e^{-i\om(v-v')}
={1 \over 4\pi } \ln\vert v - v' \vert
\label{phi2p}
\ee
across $v_H$. 
Since this equation is valid for all $v, v'$ there also exist correlations 
between $\phi_-$ and $\phi_+$. However, they become negligible for late 
Hawking quanta since these emerge from configurations which are characterized by 
frequencies $\om= \la e^{\kappa u} \gg \kappa$ and which are extremely 
localized across $v_H$, see \cite{MP2}.
These two effects follow from the asymptotic ($\kappa u \gg 1$)
relation \cite{Hawk} between the value of $u$ 
of the geodesic which originates from $v$ on ${\cal{J}}^-$
\be
V(u)-v_H \simeq e^{-\kappa u} \, .
\label{exponrel}
\ee
It is this exponential which induces both the thermal radiation at 
temperature $\kappa / 2 \pi$
and the necessity of considering trans-Planckian frequencies. 
In the absence of gravitational interactions, it also 
tells us that $\phi_-$ and $\phi_+$ are effectively two independent fields.

Our aim is now to describe how the gravitational interactions
between $\phi_-$ and $\phi_+$
modify the semi-classical description of Hawking radiation. 
The generating functional of the whole system is 
\be
Z= \int \! {\cal{D}} \phi_- \,\,  {\cal{D}} \phi_+\,\, {\cal{D}}h \,
\, e^{i[S^{(-)}_{g+h}+ S^{(+)}_{g+h}+ S_{h, g}]} \, .
\label{Z1}
\ee
In this equation, $h$ is the linear change of the metric with respect to the 
background $g$ discussed above and $S_{h, g}$ is the
action of $h$ obtained from the Einstein-Hilbert action.
$S^{(-)}_{g+h}$ and $S^{(+)}_{g+h}$ are the actions of 
$\phi_-$ and $\phi_+$ propagating in the fluctuating geometry $g+h$.

When imposing that the metric fluctuations be spherically symmetric,
$h$ is determined
by the matter stress tensor and characterized by $\psi$ and $\mu$,
two functions of $v$ and $r$\cite{Barr}. In the place of (\ref{two}), the line element in
 the fluctuating metric is
\be
ds^2 = (1 + \psi)\big[ (1 - {2M \over r } ) dv^2 + 2 dv dr \big] + r^2 d\Omega^2_2
\label{twop}
\ee
where $M = M_0 + \mu(v,r)$. 
In this metric, the matter action is 
\be
S_{g+h} = \int  \! dv dr \big[\partial_v \phi \partial_r \phi - 
\big(1 - {2M \over r} \big) {(\partial_r \phi )^2  \over 2} \big] \, .
\label{actS}
\ee
It is independent of $\psi$, thereby 
showing  an horizon induced 2D conformal invariance, {\it c.f.}  \cite{2dhili,f1}.
 
When only quadratic terms in $h$ are kept in $S_{h, g}$ 
the functional  integration over $h$ gives
\be
Z= \int \! {\cal{D}}\phi_- \, e^{iS^{(-)}_{g}}  \int  \! {\cal{D}}\phi_+ \, 
e^{iS^{(+)}_{g} + iS_{int}} \, 
\label{Z2}
\ee
where the interaction action $S_{int}$ is a quadratic form
of the total stress tensor. It thus contains 
self-interaction terms depending on $\phi_-$ or $\phi_+$ separately.
The terms concerning $\phi_-$ only have been studied in \cite{KKW} 
and lead to small effects. Those concerning $\phi_+$ are not directly relevant 
to black hole radiation since the $\phi_+$ configurations disappear through
the horizon. We shall simply assume that these interactions do not 
significantly modify (\ref{phi2p}) for $\om \simeq \kappa$, 
a weak condition for large black holes $M \gg M_{Planck}$.

$S_{int}$ also contains cross terms which couple $\phi_-$ to $\phi_+$. 
When focusing at the near horizon physics, i.e. in the semi-infinite 
strip $\vert r-2M \vert \ll 2M$, $v > 0$, the relevant metric fluctuation
is given by a mass term $\mu_+$ 
which is determined by $T_{vv}= (\partial_{v}\phi_+)^2$
\be
\mu_+(v) = \mu(v,r)\vert_{r= 2M} = \int_0^v  \! \! dv' T_{vv}(v') \, .
\label{mu1}
\ee
Using this approximation, the dominant contribution 
to $S_{int}$ is given by, see (\ref{actS}), 
\be
S_{int}= \int_0^{\infty} \!\!dr \int_0^{\infty}  \!\! dv { \mu_+(v) \over r } 
(\partial_r \phi_-)^2 \,.
\label{sint}
\ee
The reason that $ \mu_+$ gives the dominant contribution is that it
is coupled to 
$ (\partial_r \phi_-)^2 \simeq (\partial_v \phi_-)^2 /(r/2M -1)^2$ 
which diverges for $r \to 2M$. 
In brief, (\ref{sint}) represents the  $\phi_-  \phi_+$ interactions mediated by gravity.
They have been already considered in
\cite{THooft,Verl3,Casher}.
The novelty of this letter lies in their collective treatment 
when the state of $\phi_+$ is vacuum.

When using (\ref{sint}) in  (\ref{Z2})
three approximations can be considered. 
The first one is that adopted by Hawking. It simply consists in 
putting $\mu_+ =0$. Then $\phi_-$ is a free field propagating
in the background geometry $g$ and $\phi_+$ drops out 
from all matrix elements built with the operator $\phi_-$.  
In this treatment these matrix elements
are characterized by trans-Planckian frequencies 
when one of the operator approaches
the future horizon\cite{MP2}. 

The second approximation consists 
in working with the expectation value of (\ref{mu1}).
Then the mean metric change $\langle \mu_+(v) \rangle$ 
is driven by the (properly subtracted\cite{GO}) expectation value
of $T_{vv}$
\be
\langle T_{vv}(v) \rangle\vert_{r=2M} = - {  \pi  \over 12 } \big({\kappa \over 2\pi}\big)^2 \, .
\label{meanT}
\ee
This is minus the usual value of a 2D thermal flux. 
It is this negative flux which drives the black hole evaporation, see (\ref{one}).
In this mean field approximation, one simply replaces $M_0$ by  
$M_0+\langle \mu_+ \rangle$ in the former treatment. 
Therefore, the trans-Planckian problem mentioned above stays as
such 
as long as this mass loss is slow compared to $\kappa$, 
i.e. as long as $ M(v) \gg M_{Planck} $.

To solve this problem clearly requires to take into account 
the fluctuating character of the interactions between $\phi_-$ and $\phi_+$.
By proceeding formally, one would perform in (\ref{Z2}) the integration
over $\phi_+$ so as to obtain the influence functional (IF) \cite{FHibbs} 
for the  $\phi_-$ configurations on which 
we have access asymptotically.
However to perform exactly this functional integration is out of reach
since the final state of  $\phi_+$ is correlated to that of $\phi_-$. 
A convenient approximation  
consists in neglecting the entanglement.
This is a common procedure both in QFT 
where it gives the vacuum contribution, see Chapt. 9
 in \cite{FHibbs}, and in statistical mechanics 
(e.g. the {\it polaron}, Chapt. 11) when 
one neglects the modification of the environment due 
to the interactions with the system. 

In this third approximation, the IF
gives rise to a non-local action which is a sum of terms 
containing  $(\partial_r \phi_-)^2$ and kernels
given by the Wick contractions of the $\partial_v \phi_+$. 
The first term is quadratic in $(\partial_r \phi_-)^2$ and 
the kernel is the (connected) two-point function of $T_{vv}$
\be
\langle  T_{vv}(v)  T_{vv}(v') \rangle_c = {1 \over 16 \pi^2 }{1 \over (v-v')^4 }
\label{Tfluct}
\ee
evaluated in the vacuum (\ref{phi2p}).
Keeping only this term in the IF
is equivalent\cite{Verda} to work with a stochastic  
ensemble of metric fluctuations $\mu_{+}$ which obey, see (\ref{mu1}),
\ba
\langle  \mu_{+} (v)  \mu_{+} (v') \rangle &=& {1 \over 96 \pi^2 } 
{1 \over (v-v')^2 } 
\nonumber\\
&=& {1 \over 96 \pi^2 } \int_0^\infty \!\! d\om \om  \cos[\om(v-v')] \, .
\label{mf2}
\ea
This equation gives the mean properties of the spherical metric 
fluctuations driven by $\phi_+$ in its vacuum state. 

In this quadratic approximation, to exploit the near horizon conformal invariance
of (\ref{actS}), one should perform the integration over $\phi_+$ after that over 
 $\phi_-$ \cite{f3}. 
In this eikonal-like treatment, the non-linear effects induced by the metric fluctuations
are taken into account 
through the characteristics of the equation of motion of $\phi_-$ 
\be
(1 - { 2M_0 + 2 \mu_+(v) \over r }) \partial_r  \phi_- = 2 \partial_v \phi_- \, .
\label{elag}
\ee
These are nothing but the outgoing null geodesics $u(v,r)$, the non trivial 
solutions of $ds^2=0$ of 
(\ref{twop}).
The background solution is $u_0= v - 2 r^*$.
The first order change induced by $\mu_+$, $\delta u = u - u_0 $,  is 
given by \cite{Barr} 
\ba
\delta u(v)\vert_{u_0} &=& \int^\infty_{v}\!\! dv' { \mu_+(v') \over 
r(v')\vert_{u_0} - 2M_0} 
\label{17}
\ea
where $r(v)\vert_{u_0}$ is obtained by inverting $u_0(v,r)= v - 2 r^*$.
The integral is dominated by the near horizon region wherein 
$r(v)\vert_{u_0} - 2M_0 \simeq 2M_0 e^{\kappa(v-u)}$.
This dependence in $\kappa v$ will tame 
the UV content of the metric fluctuations.

To determine the effects of these fluctuations on outgoing configurations, 
we analyze the near horizon behavior of asymptotic plane waves 
representing Hawking quanta. (Notice that it also controls 
that of matrix elements of $\phi_-$ such as the Feynman Green function.)
In the absence of metric fluctuations $e^{-i \la u}$ behaves as
\be
e^{-i \la u_0(v,r)} = \theta(r-2M_0) e^{-i \la v} (r-2M_0)^{i\kappa \la} \, .
\label{19}
\ee
It vanishes for $r< 2M_0$ and 
possesses an infinite number of oscillations as $r \to 2M_0$ 
with increasing momentum $p_r= -i \partial_r$. This is 
the trans-Planckian problem. 

When considering the Feynman Green function  obtained from 
(\ref{Z2}), (\ref{sint}) and (\ref{Tfluct}) (and with one operator 
at $v,r$ and the other on ${\cal J}^+$), its behavior in $v,r$ is
determined \cite{Barr} by the ensemble average waves
\be
\langle e^{-i \la u(v,r)} \rangle \simeq e^{-i \la u_0(v,r)} 
e^{- {\la^2 \over 2}\langle \delta u(v)\delta u(v) \rangle  } \, .
\label{18}
\ee
Using (\ref{mf2}) and (\ref{17}), one obtains 
\ba
 \langle \delta u(v)\vert_{u_0}\delta u(v)\vert_{u_0} \rangle &=& \int_0^\infty \!\! {d\om
\over 12} 
{\kappa^2 \om \over \kappa^2 + \om ^2 } { 1 \over (r /2M_0 -1)^2}
\nonumber\\
&=&
 \bar \sigma_\Lambda^2 {1 \over (r /2M_0 -1)^2}
\label{res}
\ea
where the mean spread $\bar \sigma_\Lambda$ is equal to $\kappa 
\sqrt{ \ln(\Lambda / \kappa)/12 }$. 
We have have introduced the UV cut-off $\Lambda$ to define the integral over $\om$.  
Notice that $\Lambda$ is a Lorentz scalar (since it is the 
energy of an s-wave in its rest frame) and that its value is hardly relevant since 
$\bar \sigma_{\Lambda=M} = \sqrt{2} \bar \sigma_{\Lambda=1}$.

The main result of (\ref{res}) is that $\bar \sigma_\Lambda$ is not proportional to $\Lambda$
even though $\langle \mu_+^2 \rangle \simeq \Lambda^2$.
This is because the high frequencies 
($\om \gg \kappa$) are damped by the 
integration over $v$ in (\ref{17}). 
The frequencies $\om \simeq \kappa$ dominate the integral. 

Since $\langle \delta u\delta u \rangle $
diverges as $r \to 2M_0$, (\ref{18}) tells us that the correlations between 
asymptotic quanta and early configurations, 
which existed in a given background as shown in (\ref{19}),
are washed out by the metric fluctuations 
once $r - 2M_0 \simeq \bar \sigma_{\Lambda}\simeq 1/M_0 $.
The physical reason of this loss of coherence is that the state of $\phi_+$
 becomes correlated to that of $\phi_-$\cite{THooft,Verl3}. 
Nevertheless, when tracing over both $\phi_+$ and the inner configurations of 
$\phi_-$, the asymptotic properties of Hawking radiation obtain\cite{Barr}.

Finally it should be pointed out that (\ref{18}) can be 
viewed as defining a phenomenological dissipation of outgoing waves.
In this point of view, as in condensed matter\cite{Unruh81,jaco},
one ignores the interactions with the environment
and deals only with an effective 
wave propagation governed by a non-trivial dispersion relation.


In conclusion, even though we have made many simplifying assumptions, we believe that
the following results are robust. 
(A) When propagated backwards in time
towards the star's matter, outgoing quanta are scattered by the 
metric fluctuations induced by the in-falling quantum matter fields
in their vacuum state.
(B) These interactions grow so strongly near the horizon
that the quanta are completely scattered. This prevents a perturbative 
$S$-matrix description of these interactions.
(C) The in-falling vacuum fluctuations act as a
reservoir of modes. This invites to describe these interactions
in terms of stochastic metric fluctuations.
(D) Even though the spectrum of the latter contains
all frequencies (up to a UV cut-off), their impact on outgoing configurations is 
governed by frequencies $\om \simeq \kappa$. 
(E) The stationarity of vacuum (i.e. the fact that (\ref{Tfluct}) is a function 
of $v-v'$ only) leads to stationary metric fluctuations (\ref{mf2}) and 
this, combined with the stationarity of the background metric, give a mean 
spread $\bar \sigma_\Lambda$ which is independent of $v$. 

{\it Acknowledgements.} I wish to thank Robert Brout for very useful
discussions. I also thank the organizers of the 5-{\it th} Peyresq meeting where
some of the ideas presented here crystallized. 


\end{multicols}


\begin{thebibliography}{999}
\bibitem{2dhili} In the vicinity of a black hole horizon, there is an 
induced Lorentz invariance in the $u, v$ plane. 
This can be seen from the fact that the mass term and the 
contribution of the centrifugal barrier appeared multiplied by $r-2M$ 
in the 4D d'Alembertian. Indeed, the near horizon equation reduces to 
$\partial_u \partial_v \phi =0$.
\bibitem{f1} $\partial_u \partial_v \phi =0$ is valid for all $r$ 
 only when working in the geometric optic approximation.
In the exact d'Alembertian there is a potential around $r=3M$ 
which induces partial reflection, a phenomenon irrelevant for our purposes.
\bibitem{f3} By integrating over $\phi_+$ first, one obtains a nice diagramatic
representation for the near horizon interactions between $\phi_-$ and
$\phi_+$ in its vacuum. However, this non-linear field description is not easy to 
handle.
\end{thebibliography}
\end{document}